\documentclass[11pt,preprint]{aastex}
\pdfoutput=1
\usepackage{bm}
\usepackage{graphicx}
\usepackage{footnote}
\usepackage{color}
\usepackage{amsmath}
\usepackage{mathtools}
\usepackage{pdflscape}
\usepackage{multirow}

\newcommand{\mubold}{\bm{\mu}}
\newcommand\ltsima{$\; \buildrel <\over\sim \;$}
\newcommand\simlt{\lower.5ex\hbox{\ltsima}}
\newcommand\gtsima{$\; \buildrel >\over\sim \;$}
\newcommand\simgt{\lower.5ex\hbox{\gtsima}}

\newcommand\msun {M_\odot}

\shorttitle{}
\shortauthors{Bhattacharya et al}

\begin{document}

\title{MOA-2007-BLG-400 A Super-Jupiter Mass Planet Orbiting a Galactic Bulge
K-dwarf Revealed by Keck Adaptive Optics Imaging}


\author{Aparna Bhattacharya\altaffilmark{1,2}, David~P. Bennett\altaffilmark{1,2},
 Jean~Philippe~Beaulieu \altaffilmark{3, 4},  Ian~A.~Bond\altaffilmark{9},  Naoki~Koshimoto\altaffilmark{6,7}, Jessica~R.~Lu\altaffilmark{8}, Joshua~W.~Blackman\altaffilmark{4}, Aikaterini Vandorou\altaffilmark{4}, Sean K. Terry\altaffilmark{1,10}, 
 %\\ and \\ 
 Virginie~Batista\altaffilmark{3}, Jean~Baptiste~Marquette\altaffilmark{3}, Andrew~A.~Cole\altaffilmark{4}, Akihiko~Fukui\altaffilmark{11,12}, Calen~B.~Henderson\altaffilmark{13}}

\altaffiltext{1}{Code 667, NASA Goddard Space Flight Center, Greenbelt, MD 20771, USA\\
  Email: {\tt abhatta5@umd.edu}}
\altaffiltext{2}{Department of Astronomy,
    University of Maryland, College Park, MD 20742, USA}   
\altaffiltext{3}{UPMC-CNRS, UMR 7095, Institut d’Astrophysique de Paris, 98Bis Boulevard Arago, F-75014 Paris}
\altaffiltext{4}{School of Physical Sciences, University of Tasmania, Private Bag 37 Hobart, Tasmania 7001 Australia}
\altaffiltext{5}{Space Telescope Institute, 3700 San Martin Drive, Baltimore, MD 21218, USA}
%\altaffiltext{6}{Department of Astronomy, Graduate School of Science, The University of Tokyo, 7-3-1 Hongo, Bunkyo-ku, Tokyo 113-0033, Japan}
\altaffiltext{6}{Department of Astronomy, The University of Tokyo, 7-3-1 Hongo, Bunkyo-ku, Tokyo 113-0033, Japan}
\altaffiltext{7}{National Astronomical Observatory of Japan, 2-21-1 Osawa, Mitaka, Tokyo 181-8588, Japan}
\altaffiltext{8}{University of California Berkeley, Berkeley, CA}
\altaffiltext{9}{Institute of Natural and Mathematical Sciences, Massey University, Auckland 0745, New Zealand}
\altaffiltext{10} {Department of Physics, Catholic University of America, Washington, DC 20064, USA}
%\altaffiltext{10}{Okayama Astrophysical Observatory, National Astronomical Observatory of Japan, Asakuchi,719-0232 Okayama, Japan}
\altaffiltext{11}{Department of Earth and Planetary Science, Graduate School of Science, The University of Tokyo, 7-3-1 Hongo, Bunkyo-ku, Tokyo 113-0033, Japan}
\altaffiltext{12}{Instituto de Astrof\'isica de Canarias, V\'ia L\'actea s/n, E-38205 La Laguna, Tenerife, Spain}
\altaffiltext{13}{NASA Exoplanet Science Institute, IPAC/Caltech, Pasadena, California 91125, USA}

\begin{abstract}
We present Keck/NIRC2 adaptive optics imaging of planetary microlensing event MOA-2007-BLG-400 that
resolves the lens star system from the source. We find that the MOA-2007-BLG-400L planetary system consists
of a $1.71\pm 0.27 M_{\rm Jup}$ planet orbiting a $0.69\pm 0.04\msun$ K-dwarf host star at a distance 
of $6.89\pm 0.77\,$kpc from the Sun. So, this planetary system probably resides in the Galactic bulge. The 
planet-host star projected separation is only weakly constrained due to the close-wide light curve degeneracy;
the 2$\sigma$ projected separation range is 0.6--$7.2\,$AU. This host mass is at the top end of the range of
masses predicted by a standard Bayesian analysis that assumes that all stars have an equal chance of 
hosting a star of the observed mass ratio. This and the similar result for event
MOA-2013-BLG-220 suggests that more massive stars may be more likely to host planets with a mass ratio in the 
$0.002 < q < 0.004$ range that orbit beyond the snow line. These results also indicate the importance of host star mass measurements for exoplanets found by
microlensing. The microlensing survey imaging data from NASA's {\it Nancy Grace Roman Space Telescope} 
(formerly WFIRST) mission will be doing mass measurements like this for a huge number of planetary events. This host lens is the highest contrast lens-source detected in microlensing mass measurement analysis (the lens being 10$\times$ fainter than the source). We present 
an improved method of calculating photometry and astrometry uncertainties based on the Jackknife method, 
which produces more accurate errors that are $\sim$$2.5 \times$ larger than previous estimates.   
\end{abstract}

\section{Introduction}
\label{sec-intro}
Gravitational microlensing is unique in its ability to detect low mass exoplanets \citep{bennett96} beyond the 
snow line \citep{gouldloeb1992}, where the formation of giant planets is thought to be most efficient
\citep{pollack96,lissauer_araa}. Observational results from microlensing have recently been used to constrain 
the distribution of planet-to-host star mass ratios, q beyond the snow line and found a peak at roughly a Neptune 
mass \citep{suzuki16}. More recent results have used a method pioneered by \citet{sumi10} to more 
precisely measure the location of this exoplanet mass ratio function peak and to show that the
mass ratio function drops quite steeply below the peak \citep{udalski18,jung19}.

A comparison of this observed mass ratio function \citep{suzuki18} to the predictions of the core accretion theory
as modeled by population synthesis calculations
\citep{idalin04,mordasini09} found a conflict between the smooth power law mass ratio function observed
at mass ratios of $> 10^{-4}$ and the gap predicted by the runaway gas accretion process of the core
accretion theory. This predicted mass ratio gap persisted independently of whether planetary migration was
included in the population synthesis calculations. This runaway gas accretion process has long been considered
to be one of the main features of the core accretion scenario, but it has not been previously tested. However,
the development of the core accretion theory has largely focused on the formation of planets orbiting stars
of approximately solar type, while exoplanet microlensing surveys study stars ranging from about a solar
mass down to much lower masses, including late M-dwarfs and even brown dwarfs. So, it could be that
the predicted exoplanet mass gap might still be seen for a sample of solar type stars. This possibility
can be tested by measuring the masses of the exoplanet host stars found by microlensing.

A complementary view of wide orbit exoplanet demographics can be obtained through a statistical analysis
of radial velocity data. \citep{fernandes19,wittenmyer20} have recently argued that the distribution of gas giants 
peaks or flattens out at
semi-major axes of $\sim 3\,$AU for host stars of approximately solar type. While it is possible to constrain
the distribution of exoplanets by combining planet detection methods that measure very different exoplanet
attributes \citep{clanton14a,clanton14b}, we can gain a much better understanding of exoplanet occurrence
rates with better characterized exoplanet systems \citep{bennett_wide_orb}, and this is what our
high angular resolution follow-up observations will provide.

Measurements of the angular Einstein radius, $\theta_E$, and the microlensing parallax 
amplitude, $\pi_E$, can each provide mass-distance relations \citep{bennett_rev,gaudi_araa},
\begin{equation}
M_L = {c^2\over 4G} \theta_E^2 {D_S D_L\over D_S - D_L} 
       =  {c^2\over 4G}{ {{\rm AU}^2}\over{\pi_E}^2}{D_S - D_L\over D_S  D_L}  \ .
\label{eq-m_thetaE}
\end{equation}
These can be combined to yield the lens mass in an expression with no dependence on the lens or source distance,
\begin{equation}
M_L = {c^2 \theta_E {\rm AU}\over 4G \pi_E} = {\theta_E \over (8.1439\,{\rm mas})\pi_E} \msun \ .
\label{eq-m}
\end{equation}
The angular Einstein radius can be measured for most planetary microlensing events events because 
most events have finite source effects that allow the measurement of the source radius crossing time,
$t_*$. The angular Einstein radius is then given by $\theta_E = \theta_* t_E/t_*$, where 
$\theta_*$ is the angular source radius, which can be determined from the
source brightness and color \citep{kervella_dwarf,boyajian14}. As a result, the measurement of $\pi_E$
usually results in mass measurements. Unfortunately, the orbital motion of the Earth allows $\pi_E$ 
to be determined for only a relatively small subset of events that have very long durations
\citep{gaudi-ogle109,bennett2010}, long duration events with bright source stars \citep{muraki11},
and events with special lens geometries \citep{sumi16}. The microlensing parallax program using the
{\it Spitzer} space telescope at $\sim 1\,$AU from Earth has recently expanded the number of 
events with microlensing parallax measurements \citep{udalski_ogle124,street16}, but recent studies
indicates that poorly understood systematic errors in the {\it Spitzer} photometry can contaminate
some of the {\it Spitzer} $\pi_E$ measurements \citep{spitz_vs_gal,gould20,dang20}.

The method that can determine the masses of the largest number of planetary microlensing events is
the detection of the exoplanet host star as it separates from the background source star. This method uses
an additional mass-distance relation can be obtain from a theoretical or an empirical mass-luminosity relation.
The measurement of the angular separation between the lens and source stars provides the
lens-source relative proper motion, $\mu_{\rm rel}$, which can be used to determine the angular
Einstein radius, $\theta_E = \mu_{\rm rel} t_E$.  Due to the high stellar density in the fields where microlensing
events are found, it is necessary to use high angular resolution adaptive optics (AO) or 
{\it Hubble Space Telescope} (HST) observations to resolve the (possibly blended) lens and source stars
from other, unrelated stars. Unfortunately, this is not sufficient to establish a unique identification of the
lens (and planetary host) star \citep{moa310,highres}, so it is necessary to confirm that the host star is moving away
from the source star at the predicted rate \citep{ogle169,batistaogle169}.

We are conducting a systematic exoplanet microlensing event high angular resolution follow-up program
to detect and determine the masses of the exoplanet host stars with our NASA Keck Key Strategic Mission
Support (KSMS) program \citep{bennett_KSMS}, supplemented by HST observations \citep{aparna19hst}
for host stars that are most likely to be be detected with the color-dependent centroid shift method
\citep{bennett06}. This program has already revealed a number of microlens exoplanet host stars that our 
resolved from the source stars \citep{batistaogle169,van20,bennett20}, and others that are still blended
with their source stars, but show a significant elongation due to a lens-source separation
somewhat smaller than the size of the point spread function \citep{bennett07,ogle169,aparna18}.

Our follow-up program is midway through the analysis of the 30-event extended \citet{suzuki16} sample.
We have mass measurements for 11 planets, so far, with the analysis of 4 more planetary events
at an advanced stage. We have obtained upper or lower mass limits for 2 of these events, and we
have data yet to be analyzed for 10 more events. This leaves only 3 planets from the  \citet{suzuki16} 
that are not amenable to our mass measurement methods because their durations are too short for
microlensing parallax measurements and their source stars are too bright to allow the detection of the
planetary host stars. We are also beginning to expand our analysis into events from the MOA 9-year
retrospective analysis sample. This sample is expected to have about 60 planets, including planets
like MOA-bin-1 \citep{bennett12} in orbits so wide that they would not have a detectable microlensing
signal from the star, were it not for the planet. It will also include planets in binary systems \citep{bennett_ob06284}
that were excluded from the \citet{suzuki16} sample.
Several mass measurements from this sample have
already been published \citep{sumi16,ogle0124,van20}.

Our observing program is a pathfinder for the {\it Nancy Grace Roman Space Telescope} (formerly WFIRST) mission,
which is NASA's next astrophysics flagship mission, to follow the {\it James Webb Space Telescope} (JWST).
The {\it Roman} telescope \citep{WFIRST_AFTA} includes the {\it Roman Galactic Exoplanet Survey} (RGES), 
based on the Microlensing
Planet Finder concept \citep{bennett02,mpf} that will complement the {\it Kepler} mission's statistical study
of exoplanets in short period orbits \citep{borucki11,kepler2018} with a study of exoplanets in orbits 
extending from the habitable zone to infinity (i.e.\ unbound planets). The microlens exoplanets discovered by
{\it Roman} will not require follow-up observations because the {\it Roman} observations themselves will have 
high enough angular resolution to detect the lens (and planetary host) stars itself \citep{bennett07}. Our 
NASA Keck KSMS and HST observations and analysis will help us refine this mass measurement method and 
optimize the {\it Roman} exoplanet microlensing survey observing program.

The results of the analysis of the MOA-2007-BLG-400 Keck AO follow-up observations are very similar to
the results of the analysis of planetary microlensing event MOA-2013-BLG-220 by our group \citep{van20}.
These events have planet-star mass ratios in the range $2\times 10^{-3}$ to $4\times 10^{-3}$, slightly larger
than the Jupiter-Sun mass ratio of $q \approx 10^{-3}$. The measured host star masses for both events are at 
approximately the 93rd percentile of the predicted host star masses, based on a Bayesian analysis that 
assumes that every star has an equal probability to host a planet of a given mass ratio. This indicates
that this assumption is likely to be wrong and that the distribution of planets in wide orbits may depend
sensitively the host star mass as Kepler has shown to be the case for planets in short period orbits
\citep{mulders15}.

The paper is organized as follows: Section \ref{sec-model} revisits the ground based seeing limited photometry data from 2007 and re-analyzes the light curve modeling. Section \ref{sec-Followup} describes the details of our high resolution follow up observations and the reduction of the AO images. In the next section \ref{sec-Keck}, we show the process of identifying 
the host star (which is also the lens) and the source star. In section \ref{sec-murel}, 
we determine the geocentric relative lens-source proper motion from the lens and source identification and 
show that it matches with the prediction from the light curve models. Finally in sections \ref{sec-lens} and \ref{sec-discussion}, 
we discuss the exoplanet system properties and the implications of its mass and distance measurement.   

\begin{deluxetable}{cccc}
\tablecaption{Best Fit Model Parameters
                         \label{tab-mparams} }
\tablewidth{0pt}
\tablehead{
%% Use a footnote to explain numbering.
%& & & & \multicolumn{2}{c} {MCMC averages} \\
\colhead{parameter}  & \colhead{$s<1$} & \colhead{$s> 1$} & \colhead{MCMC averages} 
}  % end header.
\startdata
%                      pleb_2    plmb_2 
$t_E$ (days) & 13.516 & 13.319 & $13.46\pm 0.37$  \\   
$t_0$ (${\rm HJD}^\prime$) & 4354.5811 & 4354.5345 & $4354.5556\pm 0.0020$ \\
$u_0$ & -0.0002765 & -0.0039059 & $-0.0022\pm 0.0022$ \\
$s$ & 0.37522 & 2.66069 & $1.6 \pm 1.2$ \\
$\alpha$ (rad) & -0.79982 & -0.79880 & $-0.815\pm 0.032$ \\
$q \times 10^{3}$ & 2.1970 & 2.2027 & $2.34 \pm 0.34$  \\
$t_\ast$ (days) & 0.04711 & 0.04713 & $0.04711\pm 0.00008$ \\
$V_S$ & 20.000 & 19.982 & $19.995\pm 0.031$ \\
$I_S$ & 18.396 & 18.377 &$18.390\pm 0.030$ \\
$\mu_{\rm rel,G}\,$(mas/yr) & 8.87 & 8.87 & $8.87 \pm 0.54$ \\
%$H_S$ & 16.663 & 16.645 &$16.657\pm 0.030$ \\
fit $\chi^2$ & 3136.95 & 3136.86 & \\
\enddata
\end{deluxetable}

\section{Revisiting Photometry and Light Curve Modeling}
\label{sec-model}
\begin{figure}
\epsscale{1.0}
\plotone{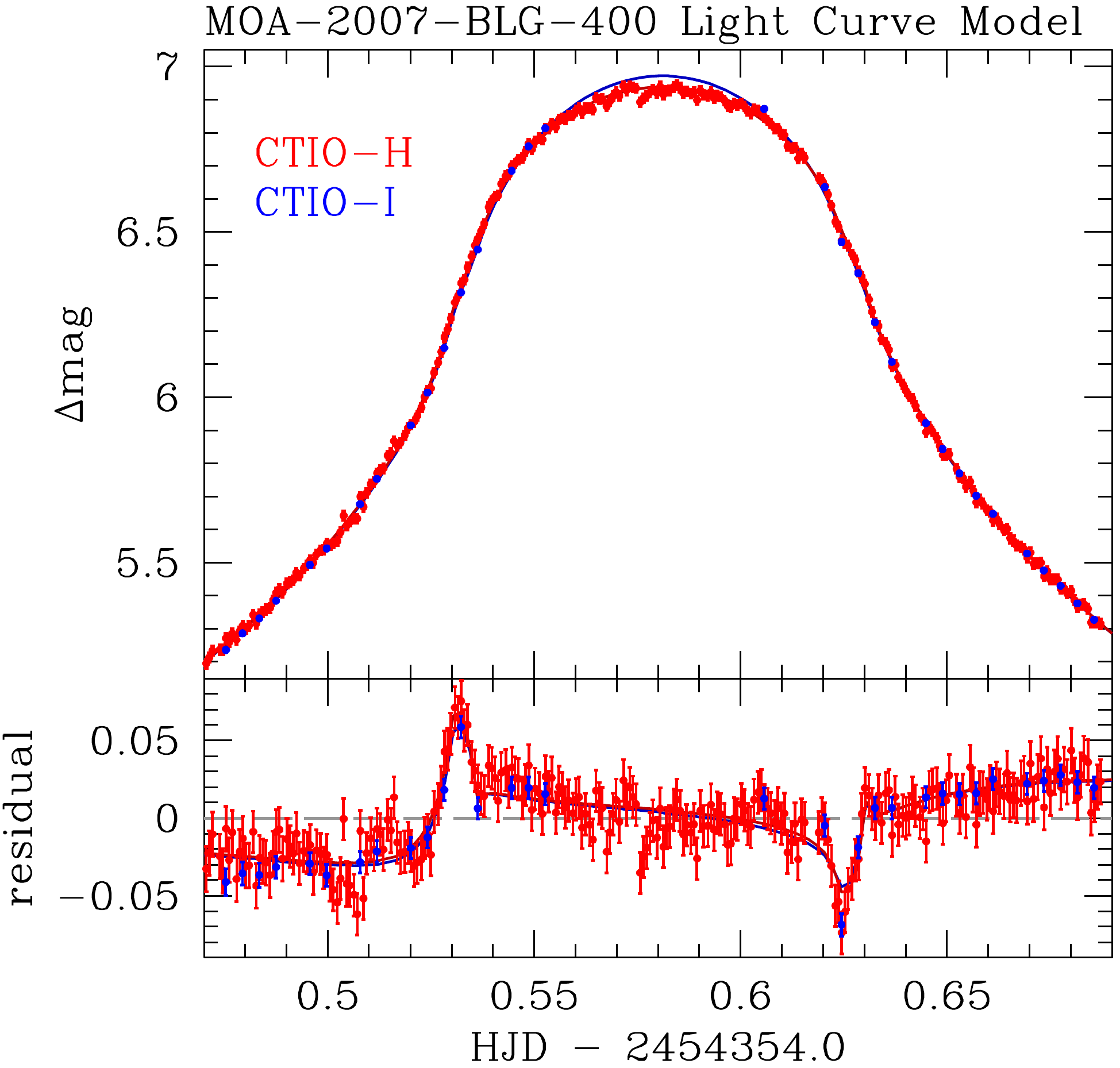}
\caption{The MOA-2007-BLG-400 light curve peak observed in the H and I bands from the SMARTS Telescope at CTIO. The top panel shows the light curve peak with different light curve for the H and I bands in red and blue. The light curve is color dependent due to limb darkening. The planetary signal reaches a maximum amplitude of 0.06 magnitudes compared to the best fit single lens light curve, but these deviations occur when the stellar limb crosses the central caustic and the magnitude changes most rapidly. Therefore, it is much easier to see the planetary signal in the residuals from the best fit single lens light curve shown in the bottom panel.}
\label{fig-Lightcurve}
\end{figure}

MOA-2007-BLG-400Lb was the eighth planet found by the microlensing method \citep{dong-moa400}, located at 
at ${\rm RA} =18$:09:41.980, ${\rm  DEC} = -29$:13:26.95, and Galactic coordinates $(l, b) = (2.3814, -4.7009)$,
There have been improvements in light curve photometry methods since the MOA-2007-BLG-400 discovery paper.
The MOA team has learned to remove some potentially serious systematic photometry errors with a detrending
method to remove systematic errors \citep{bennett12}, primarily from chromatic differential refraction. \citet{bond17}
developed a variation of the \citet{bond01} photometry method that not only included detrending but also provides
calibrated photometry. The MOA $R_{\rm moa}$ and $V$-band data, as well as the CTIO $I$ and $V$-band 
were re-reduced with the new method and  calibrated to the OGLE-III photometry database \citep{ogle3-phot}.
The CTIO $H$-band data were also reduced with the new method, but attempts to calibrate the 
CTIO $H$-band data were complicated by the unusual shape of the CTIO $H$-band PSFs. The reference frame DoPhot 
photometry \citep{dophot} we obtained for these images was unreliable, with many false ``stars" detected
due to irregularities in the $H$-band PSFs of the CTIO SMARTS Andicam images. 

Because of this improved photometry, it was necessary to redo the light curve modeling for this event.
This was done using the \citet{bennett-himag} modeling code,
and the resulting model parameters are given in Table~\ref{tab-mparams}.
The model parameters
that are in common with single lens events are the Einstein radius crossing time, $t_E$, the time, $t_0$, 
and distance, $u_0$, of closest alignment between the source and the lens center-of-mass, where $u_0$ is 
given in units of the Einstein radius. There are four additional parameters for binary lens systems: the
star-planet separation in Einstein radius units, $s$, the angle between the lens axis and the source 
trajectory, $\alpha$, the planet-star mass ratio, $q$, and the source radius crossing time, $t_*$, 
which is needed for events, like most planetary events, that have very sharp intrinsic light curve 
features that resolve the angular size of the source star. The brightness of the source star, $f_{Si}$ 
and blended stars, $f_{Bi}$ are also fit to the observed brightness for each passband, $i$, using the
formula $F_i(t) = f_{Si}A(t) + f_{Bi}$, where $A(t)$ is the magnification from the model, and $F_i(t)$
is the observed flux in the $i$the passband. Because this is a linear equation, $f_{Si}$ and $f_{Bi}$ can
be solved for exactly for each model in the Markov Chain \citep{rhie99}. For each data set
that has been calibrated, the $ f_{Si}$ values are used to determine the calibrated source brightness.

As discussed by
\citet{dong-moa400}, this event has degenerate close and wide separation light curve models with
nearly identical best fit $\chi^2$ values. The main changes
in our model parameters compared to the \citet{dong-moa400} discovery paper is that the Einstein
radius crossing time, $t_E$, has decreased by about 7\%, which is just over 2$\sigma$, and the planetary
mass ratio, $q$ has decreased by 8\%, which is $\sim 0.6\sigma$. The change in $t_E$ is due to the 
change in the MOA photometry which affects the light curve shape at low magnification, while the change
to the mass ratio, $q$, may be attributed to the combination of all data sets. Our final results are determined
from a set of six MCMC runs with a total of 263,000 light curve models. The new light curve peak is shown in Figure \ref{fig-Lightcurve}, which can be compared to Figures 1 and 2 of \citet{dong-moa400}, which show that the
new photometry and models are very similar to the \citet{dong-moa400} photometry and models.

Another improvement in our analysis is the measurement of the lens-source relative proper motion,
$\mu_{\rm rel,G} = 8.87\pm 0.54\,$mas/yr. The G suffix refers to the use of the inertial
Geocentric reference frame that moves with the
velocity of the Earth at the time of the event. This new measurement compares to $\mu_{\rm rel,G} = 8\,$mas/yr
with no error bar reported in \citet{dong-moa400}. There are several ingredients to this improvement. The improved 
CTIO photometry provides a more accurate color measurement, and the analysis of \citet{boyajian14} as
optimized for microlensing targets \citep{aparna16} provides a more accurate source radius. Finally, 
\citet{nataf13} has provided a more accurate determination of the properties of the red clump giants
that are used to determine the dust extinction in the foreground of the source. The $\mu_{\rm rel,G}$ 
prediction is important because it can be used to confirm our planetary interpretation of the
light curve \citep{ogle169,batistaogle169}.

\section{Follow up observations}
\label{sec-Followup}

The event was observed  with the Keck Adaptive Optics (AO) NIRC2 \citep{keckAO}
instrument during the  early 
morning of Aug 03, 2018 as part of our Keck NASA KSMS program. Eight dithered exposures, 
each of 30 seconds, were taken in the $K_S$ short passband with the wide camera. In this paper, from now 
on we refer to the $K_S$ band as the $K$ band. Each wide camera image covers a 1024 $\times$ 1024 square 
pixel area, and each pixel size is about $39.686 \times 39.686\,$mas$^2$. These images were flat field and dark current 
corrected using standard methods, and then stacked using the SWarp Astrometics package \citep{SWarp}. 
The details of our methods are described in \citet{batista2014}. We used aperture photometry method on these 
wide images with SExtractor software \citep{sextractor}. These wide images were used to detect 
and match fifty-seven bright isolated stars to the VVV catalog  \citep{vvv} for the calibration purposes. The 
same event was also observed with the wide camera on July 18, 2013 in the $H$ band. There are 10 wide 
camera images. These images were reduced and stacked using the same method used for the $K$ band. The 
average FWHM of this wide camera stack image is 110 mas. Fifty nine bright isolated stars were used from the
$H$ band stack image to calibrate to VVV.  Note that, in both the 
2013 and 2018 wide camera images, the lens and source were not resolved. 
As a result, we need NIRC2 narrow camera images to resolve and to identify the lens system.   

This event was also observed on Aug 03, 2018 with the Keck NIRC2 narrow camera in the $K$-band using laser guide star adaptive optics (LGSAO).  The 
main purpose of these images is to resolve the lens host star from the source star. Eleven dithered observations were taken with 
60 seconds exposures. The images were taken with a small dither of 0.7'' at a position angle (P.A.) of 0$^{\circ}$ with each frame consisting of 4 co-added 15 seconds integrations. 
%The seeing of these narrow cameraimages was $\sim$0.4-0.6''. 
The overall FWHM of these images varied from 82-98 mas.  For the reduction of these images, we used $K$-band dome flats 
taken with narrow camera on the same day as the science images. There were 5 dome flat  images with 
the lamp on and 5 more images with the lamp off, each with 60 seconds exposure time. Also at the end of 
the night, we took 10 sky images using a clear patch of sky at a (RA, Dec) of (20:29:57.71, -28:59:30.01) 
with an exposure time  of 30 seconds each. All these images were taken with the $K$ band filter. These images were used to flat field, bias subtract and 
remove bad pixels and cosmic rays from the eleven raw science images. The strehl ratio of these clean images varied
over the range 0.21-0.41. Finally 
these clean raw images were distortion corrected, differential atmospheric refraction corrected and stacked into one image. We used that for the final photometry and
astrometry analysis.

On Aug 06, 2018, we observed this event again with the NIRC2 narrow camera, but this time in the
$H$ band. We adopted an observation strategy similar to the $K$ band exposures. Seventeen dithered $H$ 
band observations were taken with 60 seconds exposures. Each exposure consisted of 3 co-added 20 second 
integrations. The $H$ band images were also taken with a small dither of 0.7'' at a position angle (P.A.) of 0$^{\circ}$. 
The overall FWHM of these images varied from 64-76 mas. We also took 10 $H$ band dome flats with the narrow camera - 5 with lamp on and the rest 5 with the lamp off. We took 15 frames of sky observations by imaging the clear patch of sky at a (RA, Dec) of (20:29:57.71, -28:59:30.01). Following the method mentioned for the $K$ band, the seventeen raw science images were cleaned using the calibration images and were stacked into one image.  The strehl ratio of these clean images varied
over the range 0.12-0.19. Both the $K$ band and $H$ band clean images were distortion corrected and stacked 
using the methods of \citet{jlu_thesis}, \citet{distortion} and \citet{refraction}. Note that, even though the 
average FWHM of the $H$ band images is smaller than that of the $K$ band images, the strehl ratios are 
significantly worse for the $H$ band images. 
This is typically the case for wavelengths shorter than the $K$ band with the current AO systems on the Keck telescopes
and other 8-10m class telescopes.

% This is probably due to the fact that AO correction for faint stars are better in $K$ band than in $H$ band.  
% Your statement above is not correct. The AO corrections are the same for faint and bright stars.

There are 1024$\times$1024 pixels in each narrow camera image with each pixel subtending
9.942 mas on each side. Since the small field of these narrow images covers only a few bright stars, 
it is difficult to directly calibrate them to VVV, so we use the wide camera $K$ and $H$ stack images that 
were already calibrated to VVV to calibrate the narrow camera images. 
This gives us the brightness calibration between the stacked narrow camera image to VVV image. The 
photometry used for the narrow camera image calibration is from DAOPHOT analysis (section \ref{sec-Keck}). 

There were also 4 $H$ band images (each of 60 seconds) of this same event taken in July 18, 2013 using 
the Keck NIRC2 narrow camera. Out of these 4 images, only 1 image has reasonably good signal to noise ratio (S/N). 
The other two images have poor S/N, probably due to the cloudy weather. In the last image, the target was outside the frame. 
Due to lack of sky images, we couldn't use the method of \citet{jlu_thesis}, \citet{distortion} and \citet{refraction} to reduce this image. 
The only good image has a FWHM of 94 mas. 
We analyzed this image directly with DAOPHOT. This analysis was used to confirm our identification of the host
(and lens) star by showing that the candidate lens star matches the motion predicted by the light curve model
between the 2013 and 2018 observations (see section \ref{sec-lens-id}).

\section{Keck Narrow Camera Image Analysis} 
\label{sec-Keck}
\subsection{2018 Narrow Data}
In this section, we use DAOPHOT \citep{Daophot} to construct a proper empirical PSF model to identify the two stars (the lens and the source) in the narrow stack images. We started our analysis with 
the 2018 Keck narrow camera images. We used the same method as \citet{aparna18} to build PSF models for both the $K$ and $H$ band narrow camera stack images. We built these 
PSF models in two stages. In the first 
stage, we ran the FIND and PHOT commands of DAOPHOT to find all the possible stars in the 
image. In second stage, we used the PICK command to build a list of bright ($K < 18.5$) isolated stars that can be 
used for constructing our empirical PSF model. Our target object was excluded from this list of PSF stars because it
is expected to consist of two stars that are not in the same position. From this list of stars, we selected the 4 nearest stars 
to the target that had sufficient brightness, and we built our PSF model from these stars. We chose only the 
nearest stars in order to avoid any effect of PSF shape variations across the image. 
We used the same PSF stars for both $K$ and $H$ band data sets. 

\begin{figure}
\epsscale{1.0}
\plotone{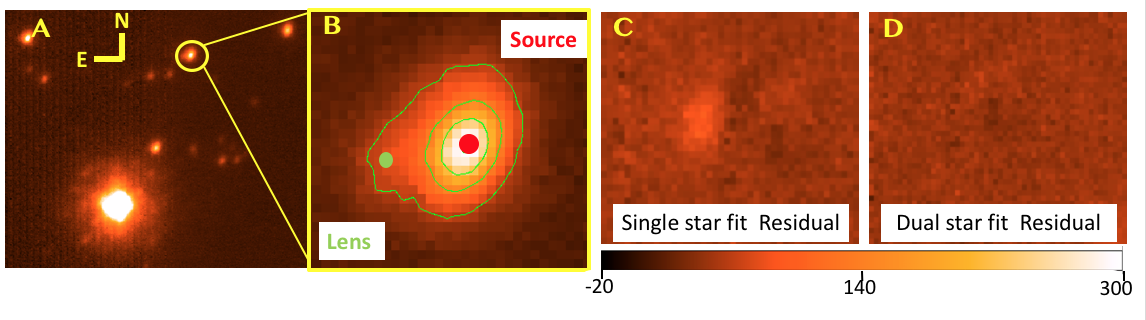}
\caption{{\it Panel A}: A 4"$\times$3" section of the stack image of 11 Keck $K$-band images, taken with the narrow camera,
with the yellow circle indicating the target.
{\it Panel B}: A 0.4"$\times$0.3' close-up the target object. The source and lens 
positions are obtained from the best fit dual star PSF model. {\it Panel C}: The residual image after subtracting 
the best fit single star PSF model. The residual shows a clear indication of the presence of an additional star East of the target. {\it Panel D}: The residual image after 
subtracting the best fit dual star PSF model. This shows a smooth residual, confirming the second star at 
a separation of $\sim$96 mas. (C) and (D) use the same photometry scale.}
\label{fig-keck}
\end{figure}

 Once we have built the PSF model, we fit all the stars with this model.  
%we carefully checked the residual image that has
%all the identified stars subtracted. We noticed that the residuals of all the single stars looked different in different parts of the image. This indicates the PSF varies over the field, an effect that is particularly pronounced for the $H$ band stack image. So, we reconstructed our empirical PSF model using only 4 bright, isolated stars near the target object. This was our final PSF model. We ran the PSF fitting again on the field 
%with this new PSF model. 
This step produced the single star residual fit for the target that is shown in Figure \ref{fig-keck}C. Inspection of this residual image from the single star fit 
indicates that there is an additional star near the target object. So, we tried fitting the region of the target object with a two-star model. The two star fits produced a smoother residual than the single star fit, as shown in Figure \ref{fig-keck}. The results of these dual star fits are given in Tables \ref{tab-Keck_phot} and \ref{tab-murel}. Both the single star and dual star fits were done using the Newton-Raphson method of standard DAOPHOT. 
As we discuss in Section~\ref{sec-keck-inter}, we identify the brighter of the two stars as the source for the
MOA-2007-BLG-400 microlensing event, and the fainter star to the East as the lens and planetary host star. The
single high quality 2013 NIRC2 narrow camera image aids in the identification of the lens star.

%We noticed that the residuals of the dual star fit in both passbands are not as clean as some of our previous analyses \citep{batistaogle169, aparna18,van20}. However, most of the single bright stars in the image have similar residual as the final two star fits. Hence, a possible reason for this could be the PSF model built from fainter stars is not sufficiently good for the bright star. A more robust way to deal with this problem is to build separate PSF models for the bright star and the faint star. {\bf [This discussion is strange. Generally, the PSF shape should not depend on brightness, unless you are close to saturation where the detector response can become non-linear. A more plausible reason for the larger residuals for brighter stars is simply that they are brighter so that their residuals are above the noise. A more likely scenario is simply that DAOPHOT has failed to construct an imperfect PSF model, which leaves significant residuals for the brighter stars.]} However, in the current version of DAOPHOT, there is no way to do dual star fits with two different PSF models simultaneously.   

\subsection{Error Bars from the Jackknife Method \label{sec-jackknife}}
We have developed a new method of error bar determination using 
the Jackknife method \citep{quenouille1949,quenouille1956, tukey1958, tierney}. This method is specifically helpful to remove the bias and measure the variance due to the 
PSF variations in the individual images. In this method, if there are N clean images, then N new stack images are 
built by excluding one of the N images from each new stack. Hence, each of these N stacks consists of N-1 images. 
Then, these N images are analyzed with DAOPHOT to build empirical PSFs for each stack image. The 4 nearby stars
used to build the empirical PSF model for the full stack of N images are used to build these empirical PSFs 
for these jackknife stack images of N-1 images. Next, the target in each image is fitted with the dual star PSF models. We built an automated code that runs the image reduction method from \citet{distortion} and \citet{refraction} and DAOPHOT routines to make these N combinations. Once we have these N combinations, we perform statistics of a parameter on these N stacked images made from N-1 images instead of 1 stacked image made from N images. The standard error of a parameter $x$ in Jackknife is given by: 
\begin{equation}
\label{eq-jackknife}
SE(x) = \sqrt{\frac{N-1}{N} \sum (x_{i} - \bar{x})^{2}} \ .
\end{equation}  
The $x_{i}$ represents the value of the parameter measured in each of the combined image and $\bar{x}$ represents the mean of the parameter $x$ from all the N stacked images. This Equation \ref{eq-jackknife} is the same 
formula as the sample mean error, except that it is multiplied by $\sqrt{N-1}$.

\begin{deluxetable}{ccccc}
\tablecaption{Measured Source and Lens Photometry and Astrometry from 2018 data\label{tab-Keck_phot}}
\tablewidth{0pt}
\tablehead{\colhead{Passband}& \multicolumn{2}{c}{Calibrated Magnitudes}&\multicolumn{2}{c}{Separation(mas)} \\
&Source &Lens& East& North}
\startdata
Keck $K$ & $16.43 \pm 0.04$ & $18.93 \pm 0.08$&$93.3 \pm 1.6$ & $-22.8 \pm 1.9 $ \\
Keck $H$ & $16.58 \pm 0.04$& $19.08 \pm 0.11$& $91.8 \pm 2.5$ & $-27.9 \pm 2.4$ \\
\enddata\\
\vspace{0.15cm}
The separation was measured 10.894 years and 10.903 years after the peak of the event in the 
$K$ and $H$ bands, respectively.
% We should report the time interval to better than 1% precision because the measurement is better than 2%.
\end{deluxetable}

We ran our automated code on $K$ and $H$ band images to build 11 $K$-band stacks of 10 combined images 
and 17 $H$-band stacks of 16 combined images, respectively. To build these stacks, our image reduction pipeline chooses the image with lowest FWHM as the reference image. However, there was one stack in both $K$ and $H$ passbands where this reference image is removed. In that case, the image with the lowest FWHM among that sample was automatically selected as the reference image. For the stack images that were built with the same reference image, the star coordinates were similar and there was no need to run FIND and PICK commands in DAOPHOT. For the stack image that had a different reference frame, the star pixel coordinates were shifted and we had to double-check by eye that indeed the same PSF stars were selected.  Once the stack images were analyzed using DAOPHOT to build empirical PSFs and fit dual star models, the error bars on lens-source separations and fluxes were calculated following Equation \ref{eq-jackknife}. We noticed that the $H$ band residual of the fits were not as smooth as the $K$ band fit. The $H$ band $\chi^2$ for the dual star fit is 4.39 times higher than the $K$ band $\chi^2$ for the same size of fitting box. As a result, we rescale the $H$ band error bars with multiplying them by $\sqrt{4.39}= 2.09$.  These error bars are reported in Table \ref{tab-Keck_phot}. The error bars in Table \ref{tab-murel} 
are based on the error bars in Table \ref{tab-Keck_phot} because the Heliocentric lens-source relative proper motion, 
$\mu_{\rm rel,H}$, is proportional to the lens-source separation.
  
\begin{figure}
\epsscale{0.9}
\plotone{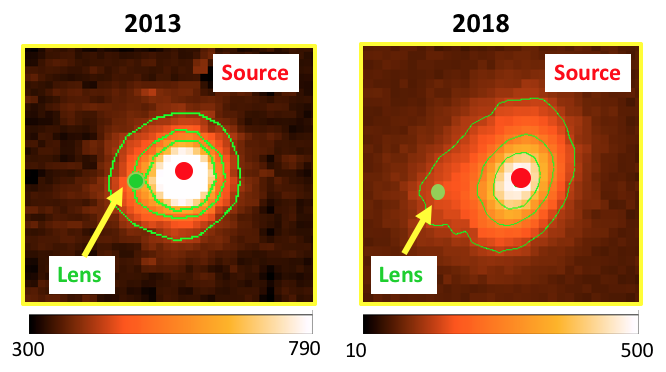}
\caption{These images show the positions for the stars that we identify as the source and lens in a 2013 $H$ band
image (left) and a 2018 $K$ band stack image. The lens star is $59.5 \pm 10.4\,$mas East and $-4.5\pm 9.0\,$mas North
of the source in the 2013 $H$ band image.
A comparison with a combined 2018 Keck $K$ image is shown on the right, where the lens is 
measured to be $93.3 \pm 1.6$ mas East and $-22.8 \pm 1.9$ mas North of the source. 
Both images are presented with East to the left, and North up. The Keck 2018 image is a combined image 
based on 11 frames, whereas 2013 image is our single high quality image from 2013. 
As discussed in section \ref{sec-lens-id}, the 2013 lens-source relative proper motion is consistent with the 2018 lens-source relative proper motion  
at  $< 1 \sigma$ in both the East and North directions.}
\label{fig-keck_2013_2018}
\end{figure}   

The running time for this software on 11 images was about 1 hour, and 17 image took about 1.5 hours on 
a quad core i7 CPU with 16 GB RAM. This shows that this method is not too time consuming and it can be easily 
run for 70 images over a few days. Since our largest data set for a single event is $\approx 70$ images, 
this indicates that this jackknife method can be used for all of the events we observe in this program.  
  
\subsection{2013 Narrow Data}\label{sec-2013narrow}
We ran the DAOPHOT analysis on the single high quality 2013 $H$ band image using the same method as 
for 2018 data. We used 4 PSF stars to build the empirical PSF model and run a dual star fit. With only a single
image, we obviously cannot use the jackknife method to estimate the position error bars. However, we expect 
the same image PSF variance as we saw in the 2018 $H$ band data. So, we take the 2018 $H$ band position
errors and multiply them by $\sqrt{17}$ to account for the 17 images contributing to the 2018 result compared to the
single 2013 image. This yields an offset between the faint and bright stars of 
$59.5 \pm 10.4\,$mas East and $-4.5 \pm 9.0\,$mas North.
The positions of the stars that we identify as the lens and source are shown in Figure \ref{fig-keck_2013_2018}
for both the 2013 and 2018 measurements.

\section{Interpreting Our Keck Results \label{sec-keck-inter}}     

In previous sections, we have referred to the Keck images of the source and lens (and planetary host) stars.
Now, we discuss the analysis that shows that our identifications of the source and lens stars are correct. 
For a high magnification event, like MOA-2007-BLG-400, it is possible to determine the position of the 
the lens and source stars at the time of the event to a precision of better than $20\,$mas based on the 
MOA difference images when the source was highly magnified. This precision might be slightly degraded
by the coordinate transformation between the MOA images and the Keck NIRC2 wide camera images,
but this analysis, based on the 2013 $H$ band wide images uniquely identifies the target star. This target star is shown by the
yellow circle in Figure~\ref{fig-keck}A.

\subsection{Confirming the Source Star Identification}
 Our reanalysis of the light curve model of this event 
(with improved MOA and CTIO photometry) showed the best fit model (wide) gives the source brightness $I_S = 18.38 \pm {0.03}$ and $V_S = 19.98 \pm {0.03}$. The average extinction and reddening of the red clump stars within $2.5^\prime$ of the target are $A_I = 0.92$ and $E(V-I)= 0.75$, as calculated from the OGLE-III photometry catalog \citep{ogle3-phot}. 
This means the source color is $(V-I)_S = 1.61 \pm {0.04} $, and 
the extinction corrected color is $(V-I)_{S,0} = 0.86 \pm {0.04} $. From the color-color relation 
of \citet{kenyon_hartmann}, 
$(V-I)_{S,0}=0.86$ corresponds to the dereddened colors of $(I-H)_{S,0}=1.03$ and $(I-K)_{S,0}=1.11$. 
So the extinction corrected source brightnesses in $K$ and $H$ bands are $K_{S,0} = 16.35 \pm {0.06}$ and $H_{S,0} = 16.43 \pm {0.06}$, including 
a 5$\%$ uncertainty in the color-color relation. This implies the dereddened color $(H-K)_{S,0}=0.08$ which is consistent with estimates from \citet{kenyon_hartmann}. From \citet{cardelli}, the extinctions 
in the $K$ and $H$ bands are $A_K = 0.13$ and $A_H = 0.21$ at 6.9 kpc (see section \ref{sec-lens}). 
Hence, the $K$ band and $H$ band 
calibrated magnitudes of the source are $K_S = 16.48 \pm 0.06$ and $H_S = 16.64 \pm {0.06}$.     

\begin{deluxetable}{ccccc}
\tablecaption{Measured Lens-Source Relative Proper Motion\label{tab-murel}}
\tablewidth{0pt}
\tablehead{\colhead{Passband}&
\multicolumn{2}{c}{$\mathbf{\mu}_{\rm rel,H}$(mas/yr)} &
\multicolumn{2}{c}{$\mathbf{\mu}_{\rm rel,H}$(mas/yr)}\\
&East&North& galactic-$l$ & galactic-$b$ }
\startdata
 Keck $K$ &  $8.56 \pm 0.15$ & $-2.09 \pm 0.17$ & $2.26 \pm 0.17$ & $-8.52\pm 0.16$ \\
 Keck $H$  & $8.43 \pm 0.23$ & $-2.56\pm 0.19$ & $1.79\pm 0.21$ & $-8.63\pm 0.22$ \\
Mean &$8.52\pm0.13$ &$-2.29\pm 0.13$ &  $2.07\pm 0.13$ & $-8.55\pm 0.13$ \\
\enddata
\end{deluxetable}

The dual star fits imply that the two stars located at the position of the target are completely separated. 
The best dual star fit yielded two stars with calibrated $K$ magnitudes of $16.43 \pm 0.04$ and $18.93 \pm 0.08$. 
The $H$ magnitudes for those same stars are $16.58 \pm 0.04$ and $19.08 \pm 0.11$. The error bars are 
calculated using the jackknife method, as discussed in subsection \ref{sec-jackknife}. The brighter star from 
these dual star fits matches the source magnitudes derived from the light curve modeling, i.e, 
$K_{\rm source} =16.48 \pm 0.06$ and $H_{\rm source} =16.64 \pm {0.06}$. 
Hence, we identify the brighter star as the source star and the fainter star as the potential lens star.
 
\subsection{Confirmation of the Host Star Identification}
\label{sec-lens-id}

The 2013 analysis indicates that the $H$ band magnitudes of the source and the fainter stars are $16.61 \pm 0.04$ 
and $19.01 \pm 0.11$, respectively. This is consistent with the 2018 $H$ band 
magnitudes of $16.58 \pm 0.04$ and $19.08 \pm 0.11$, respectively.
The source and the fainter star are separated by slightly more than the FWHM of the 2018 $H$ and $K$ band images,
and the measured separation can be used to compute the lens-source relative proper motion, $\mu_{\rm rel}$, 
which can be compared with the $\mu_{\rm rel}$ prediction from the light curve. However, this issue is complicated
by the fact that the lens-source separation measurements determine the relative proper motion in a Heliocentric
frame, $\mu_{\rm rel,H}$, while the light curve measures the relative proper motion in an intertial Geocentric reference
frame, $\mu_{\rm rel,G}$, that moves with the Earth's velocity at the time of the event. The relationship between
these reference frames given by equation \ref{eq-mu_helio}, and discussed in detail in Section~\ref{sec-murel},
but for most events, and especially for events like MOA-2007-BLG-400, where a distant lens is favored, we
have $\mu_{\rm rel,H}\approx \mu_{\rm rel,G}$ to a good approximation. 

The simplest $\mu_{\rm rel}$ comparison is to compare the measured separations in the 2018 and 2013 Keck 
observations. As reported in Section~\ref{sec-2013narrow}, we find a lens-source separation of 
$59.5 \pm 10.4\,$mas to East and $-4.5 \pm 9.0\,$mas to North. Dividing by the 5.83 year interval between
the microlensing event peak and the 2013 Keck, observations, we find
$(\mu_{\rm rel,H,E}, \mu_{\rm rel,H,N}) = (10.2 \pm 1.8, -0.8 \pm 1.5)\,$mas/yr. These values are each within
1$\sigma$ of the mean $\mu_{\rm rel,H}$ values reported in Table~\ref{tab-murel} and discussed in more 
detail in Section~\ref{sec-murel}. This consistency implies that both the 2013 and 2018 measurements are
consistent with the same lens-source relative motion and with the lens and source being in the same position 
at the time of the microlensing event.

In order to compare to the light curve prediction of $\mu_{\rm rel,G} = 8.87\pm 0.54\,$mas/yr, we must use 
equation~\ref{eq-mu_helio} to convert between the Geocentric and Heliocentric coordinate systems. This
requires knowledge of the source and lens distances, and we do this comparison inside our Bayesian 
analysis, presented in Section~\ref{sec-lens}, which combines the Markov chain of light curve models
with the constraints from the Keck observations and the Galactic models. The inclusion of the Keck 
proper motion constraints in these calculations modifies the light 
curve prediction of $\mu_{\rm rel,G} = 8.87\pm 0.54\,$mas/yr to
$\mu_{\rm rel,G} = 8.77 \pm 0.13\,$mas/yr. This is with 0.2$\sigma$ of the light curve prediction, and 
reduces the final uncertainty by a factor of more than 4. So, the Keck measurement obviously matches
the light curve prediction.

These analyses completely rule out a possible source companion as the source
of the flux that we attribute to the lens star. The motion between 2013 and 2018 is inconsistent with
a star bound to the source. An unrelated star in the bulge would have to mimic the proper motion of
the lens star, and the probability of this is $< 0.06\,$\% according to an analysis using the method
of \citet{highres}. There is also the possibility that we have detected a binary companion to the lens instead
of the lens star. This possibility is severely constrained because of the sensitivity of this high magnification event to
binaries with projected separation $\simlt 200\,$AU. However, it is only a small
fraction of possible wide separation binary companions that will have a separation from the source that is consistent
with the light curve prediction for the lens. An analysis following the method of  \citet{highres}, indicates a 
probability of $<0.1\,$\% that the star we identify as the lens star is actually a binary companion to the lens
star.
 
\section{Determination of Relative Lens-Source Proper Motion}
\label{sec-murel}

Our high resolution observations were taken $\sim$10.9 years after the microlensing event magnification peak. If these images were taken exactly 11 years after the microlensing magnification, then we would not need to consider the relative motion of the Earth with respect to the Sun when determining the lens-source relative proper motion.
The relative lens-source motion between the time of the event magnification peak and the Keck observations
moved the Earth $\sim 0.6\,$AU with respect to the Heliocentric frame. With a lens distance of $6.89\,$kpc 
(see section \ref{sec-lens}), this implies an angle of $\sim 0.09\,$mas, which is $\ll 1\sigma$ of the 
separation measurement uncertainties given in Table~\ref{tab-murel}. Therefore, we are safe in assuming that
our Keck measurement gives the relative proper motion in the Heliocentric reference frame, $\bm{\mu}_{\rm rel,H}$.

At the time of peak magnification, the separation between lens and source was 
$\sim |u_0\theta_E| \sim 0.001\,$mas. 
Hence, by dividing the measured separation by the time interval 
of 10.894 years (for $K$ band or 10.903 years for $H$ band), we obtain the heliocentric lens-source 
relative proper motion, $\mubold_{\rm rel,H}$. A comparison of these values from our independent dual star
fits for the $H$ and $K$ bands are shown in Table \ref{tab-murel} with 
error bars are estimated from jackknife method. In Galactic coordinates, the mean $\mubold_{\rm rel,H}$ components are 
$\mu_{{\rm rel,H},l} = 2.07 \pm 0.13$ mas/yr and  
$\mu_{{\rm rel,H},b} = -8.55\pm 0.13$ mas/yr, with an amplitude
of $\mu_{\rm rel,H} = 8.80\pm 0.18$ mas/yr at an angle of $\sim -74^\circ$ from the direction of Galactic disc rotation. 
The dispersion in the motion of stars in the bar shaped bulge at the lens distance of 
$D_L \approx 6.97\,$kpc (as presented in section \ref{sec-lens}) is about $\sim 2.5\,$mas/yr in each direction. 
The source is also in the bulge at about $\sim 8\,$kpc, where a similar dispersion in the motion of stars is expected. 
The relative proper motion is the difference of two proper motions, so the average difference in proper motion is
the quadrature sum of four  $\sim 2.5\,$mas/yr values or $\sim 5\,$mas/yr. However the microlensing rate 
is proportional to $\mu_{\rm rel}$, so the average $\mu_{\rm rel}$ is greater than $> 5\,$mas/yr. So, our
measured $\mubold_{\rm rel,H}$ value is only slightly higher than the typical value for bulge-bulge lensing events.

Our light curve models were done in a geocentric reference frame that differs from the heliocentric
frame by the instantaneous velocity of the Earth at the time of peak magnification, because the 
light curve parameters can be determined most precisely in this frame. However, this also means that
the lens-source relative proper motion that we measure with follow-up observations is not in the
same reference frame as the light curve parameters. This is an important issue because, as we
show below (see section \ref{sec-lens}), the measured relative proper motion can be combined with brightness of
the source star determine the mass of the lens system. The relation between the relative
proper motions in the heliocentric and geocentric coordinate systems are given by \citet{dong-moa400}:
%{\bf [\citep{dong-moa400} is the wrong reference. Find the right one.]}
\begin{equation}
\bm{\mu}_{\rm rel,H} = \bm{\mu}_{\rm rel,G} + \frac{{\bm v}_{\oplus} \pi_{\rm rel}}{\rm AU}  \ ,
\label{eq-mu_helio}
\end{equation}
where ${\bm v}_{\oplus}$ is the projected velocity of the earth relative to the sun (perpendicular to the 
line-of-sight) at the time of peak magnification. The projected velocity for MOA-2007-BLG-400 is
${{\bm v}_{\oplus}}_{\rm E, N}$ = (6.9329, -2.8005) km/sec = (1.46, -0.59) AU/yr at the peak of the microlensing. The relative parallax is 
defined as $\pi_{\rm rel} \equiv  (1/D_L - 1/D_S)$, where $D_L$ and $D_S$ are lens and source distances. Hence the Equation \ref{eq-mu_helio} can written as:
\begin{equation}
\bm{\mu}_{\rm rel,G} = \bm{\mu}_{\rm rel,H} - (1.46, -0.59 )\times (1/D_L - 1/D_S) 
\label{eq-mu}
\end{equation}
Since $\bm{\mu}_{\rm rel,H}$ is already measured in Table \ref{tab-murel}, Equation 4 yields the geocentric relative proper motion, $\bm{\mu}_{\rm rel,G}$ as a function of the lens distance. Now at each possible lens distance, 
we can use the $\mu_{\rm rel,G}$ value from equation~\ref{eq-mu} to determine the angular 
Einstein radius, $\theta_E = \mu_{\rm rel,G} t_E$. Since we already know the $\theta_E$ value from the light curve models, we can use that here to constrain the lens distance and relative proper motion. Using this method, we determined the vector
relative proper motion to be  $\bm{\mu}_{{\rm rel,G}}({E,N}) = (8.49 \pm 0.14, -2.28 \pm 0.14)$ and the magnitude
of this vector is $\mu_{\rm rel,G} = 8.79 \pm 0.18$. 
This value is consistent with the predicted $\mu_{\rm rel,G} = 8.77 \pm 0.13$ mas/yr from the light curve models.

\section{Lens Properties}
\label{sec-lens}
In order to obtain good sampling of light curves that are consistent with our photometric 
constraints and astrometry, we apply the following constraints, along with Galactic model constraints
when summing over our light curve modeling MCMC results to determine the final parameters.
The constraints are: $\mu_{\rm rel,H,l}$ and $\mu_{\rm rel,H,b}$ are constrained to have the values
and error bars from the bottom row of Table \ref{tab-murel}, and the lens magnitudes constrained to 
be $K_L = 18.93 \pm 0.08$ and $H_L = 19.08 \pm 0.11$. 
The $\mu_{\rm rel,H}$ constraints are applied to the Galactic model and the lens magnitude
constraints are applied when combining the MCMC light curve model results with the Galactic model.
The lens magnitude constraints require the use of a mass-luminosity relation. We built
an empirical mass luminosity relation following the method presented in \citet{bennett_moa291}.
This relation is a combination of mass-luminosity relations for different mass ranges. 
For $M_L \geq 0.66\,\msun$, $0.54\,\msun\geq M_L \geq 0.12\,\msun $, and 
$0.10 \,\msun \geq M_L \geq 0.07\,\msun$, we use the relations of \citet{henry93}, \citet{delfosse00},
and \citet{henry99}, respectively. In between these
mass ranges, we linearly interpolate between the two relations used on the
boundaries. That is, we interpolate between the \citet{henry93} and the \citet{delfosse00}
relations for $0.66\,\msun > M_L > 0.54\,\msun$, and we interpolate between the
\citet{delfosse00} and \citet{henry99} relations for $0.12\,\msun > M_L > 0.10\,\msun$. When using
these relations we assume a 0.05 magnitude uncertainty.

For the mass-luminosity relations, we must also consider the foreground extinction.
At a Galactic latitude of $ b = -4.7009^\circ$, most of the dust is likely to be in the foreground of
the lens unless it is very close to us. We quantify this with a relation relating the extinction 
if the foreground of the lens to the extinction in the foreground of the source.
Assuming a dust scale height of $h_{\rm dust} = 0.10\pm 0.02\,$kpc, we have
\begin{equation}
A_{i,L} = {1-e^{-|D_L(\sin b)/h_{\rm dust}|}\over 1-e^{-|D_S (\sin b)/h_{\rm dust}|}} A_{i,S} \ ,
\label{eq-A_L}
\end{equation}
where the index $i$ refers to the passband: $V$, $I$, $H$, or $K$.

These dereddened magnitudes can be used to determine the angular source radius,
$\theta_*$. With the source magnitudes that we have measured, the most precise determination
of $\theta_*$ comes from the $(V-I),I$ relation. We use
\begin{equation}
\log_{10}\left[2\theta_*/(1 {\rm \mu as})\right] = 0.501414 + 0.419685\,(V-I)_{s0} -0.2\,I_{s0} \ ,
\label{eq-thetaS}
\end{equation}
which comes from the \citet{boyajian14} analysis, but with the color range optimized for the 
needs of microlensing surveys \citep{aparna16}.

\begin{deluxetable}{cccc}
\tablecaption{Measurement of Planetary System Parameters from the Lens Flux Constraints\label{tab-params-histogram}}
\tablewidth{0pt}
\tablehead{\colhead{parameter}&\colhead{units}&\colhead{values \& RMS}&\colhead{2-$\sigma$ range}}
\startdata
Angular Einstein Radius, $\theta_E$&mas&$0.32\pm 0.01$&0.30--0.34 \\
Geocentric lens-source relative proper motion, $\mu_{\rm rel, G}$&mas/yr&$8.77\pm  0.13$& 8.50--9.03\\
Host star mass, $M_{\rm host}$&${\msun}$&$0.69\pm 0.04$ & 0.61--0.78\\
Planet mass, $m_p$&$M_{\rm Jup}$& $1.71\pm 0.27$& 1.25--2.32\\
Host star-planet 2D separation, $a_{\perp}$&AU&$3.5\pm 2.7$& 0.6--7.4\\
Host star-planet 3D separation, $a_{3D}$&AU&$4.9^{+3.3}_{-4.1}$& 0.7--18.3\\
Lens distance, $D_L$&kpc &$6.89\pm 0.77$& 5.57--8.48\\
%Lens magnitude, $K_L$& &$18.78\pm 0.11$& 18.57--18.99\\
%Lens magnitude, $H_L$& &$19.02\pm 0.12$& 18.81--18.99\\
Lens magnitude, $I_L$& &$21.24\pm 0.13$& 20.92--21.77\\
Lens magnitude, $V_L$& &$23.46\pm 0.25$& 22.91--24.29\\
\enddata
\end{deluxetable}

\begin{figure}
\epsscale{1.0}
\plotone{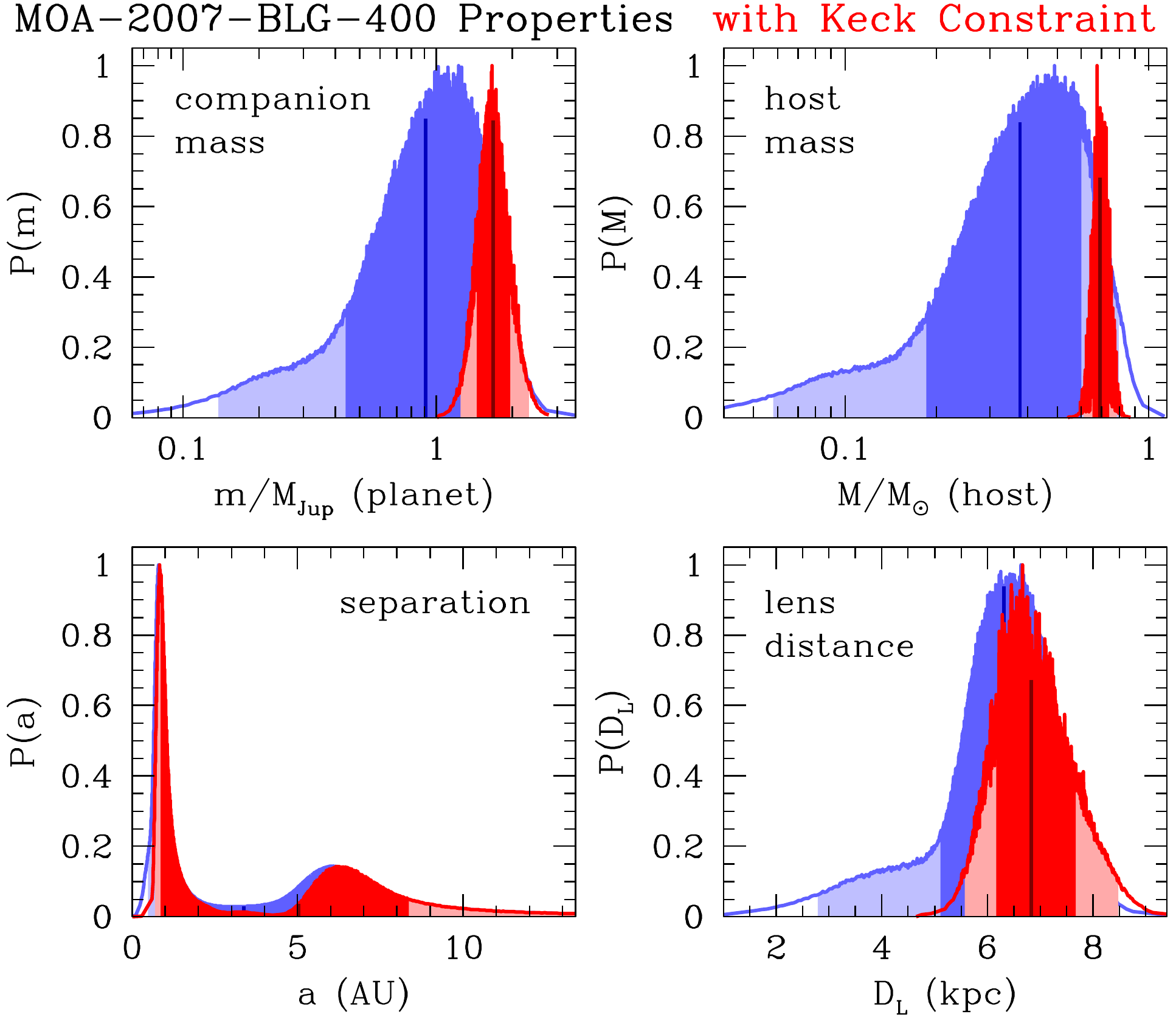}
\caption{The Bayesian posterior probability distributions for the planetary companion mass, host mass, 
their separation and the distance to the lens system are shown with only light curve constraints in blue 
and with the additional constraints from our Keck and HST follow-up observations in red.
The central 68.3$\%$ of the distributions are shaded in darker colors (dark red and dark blue) and the 
remaining central 95.4$\%$ of the distributions are shaded in lighter colors. The vertical black line marks 
the median of the probability distribution of the respective parameters.}
\label{fig-lens}
\end{figure}

We apply the $H$ and $K$-band mass-luminosity relations to each of the models in our Markov Chains
using the mass determined by the first expression of equation~\ref{eq-m_thetaE}, using the $\theta_E$
value determined from $\theta_E = \theta_* t_E/t_*$, where $t_E$ and $t_*$ are light curve parameters
given in Table~\ref{tab-mparams}. We can then use the Keck $H$ and $K$-band measurements of the
lens star brightness from Table~\ref{tab-Keck_phot} to constrain the lens brightness including both the
observational uncertainties listed in  Table~\ref{tab-Keck_phot} and the 0.05 mag theoretical uncertainty
that we assume for our empirical $H$ and $K$-band mass-luminosity relations.
To solve for the planetary system parameters, we sum over our MCMC results
using the Galactic model employed by \citet{bennett14} as a prior, weighted by the 
microlensing rate and the measured $\mubold_{\rm rel,H}$ values given in Table~\ref{tab-murel}. 

The results of our final sum over the Markov Chain light curve models are given in 
Table~\ref{tab-params-histogram} and Figure \ref{fig-lens}. This table gives the mean and RMS
uncertainty plus the central 95.4\% confidence interval range for each parameter except the 
3D separation, $a_{3D}$, where we give the median and the central 68.3\% confidence interval
instead of the mean and RMS. The lens flux and the $\mubold_{\rm rel,H}$ measurements   
exclude most of the masses and distances for this planetary system that were compatible with 
Bayesian analysis and MCMC light curve models without any $\bm{\mu}_{\rm rel,H}$ or lens brightness 
constraints. The host mass is measured to be $M_{\rm host} = 0.69\pm 0.04\msun$, a K dwarf star, orbited by a 
super-Jupiter mass planet,
$M_{\rm Jup} = 1.71\pm 0.27$ at a projected separation of $a_{\perp} =3.5\pm 2.7\,$AU. The separation
distribution is bimodal with the close and wide solutions yielding $a_{\perp} = 0.79\pm 0.10\,$AU and
$a_{\perp} =5.9\pm 0.7\,$AU, respectively. This analysis also
implies a lens system distance of $D_L = 6.97\pm 0.77$ kpc. These results show that the planet is
slightly less than twice the mass of Jupiter orbiting a K-dwarf that is very likely to be in the bulge. 

Figure \ref{fig-lens} shows the posterior distributions of the planet and host masses, their 3-dimensional
separation (assuming a random orientation) and the distance to the planetary lens system. The blue
histograms in this figure show the results based on only the $\theta_E = \theta_* t_E/t_*$ measurement
from the light curve, and the Galactic prior. This calculation also makes the assumption that possible host 
stars of any mass are equally likely to host a planet of the observed mass ratio, $q = (2.34\pm 0.34)\times 10^{-3}$
and separation, $s = 0.365 \pm 0.024$ or $s = 2.72 \pm 0.14$ (for the degenerate close and wide models).
The red histograms are the results after including the constraints from our Keck $H$ and $K$-band 
AO images. These show that the host mass is near the
top of the range predicted by the analysis prior to the Keck constraints. In fact, the host mass is in the 
93rd percentile of the predicted distribution. 

The Keck observations have constrained the host star $K$ and $H$-band magnitudes, as indicated in 
Table~\ref{tab-Keck_phot}, and Table~\ref{tab-params-histogram} also gives the inferred host star
$V$ and $I$-band magnitudes. A comparison of these results with the data used for Figure 7 of
the \citet{dong-moa400} reveals that the measured value of the host star magnitude, $H_L  = 19.08 \pm 0.11$,
is significantly brighter than the 2$\sigma$ lower limit, $H_L > 19.47$ from the discovery paper. The 
measured 2$\sigma$ upper limit, $H_L < 19.074$ has a probability of only 0.2\% according to the
analysis of \citet{dong-moa400}. Similarly, our 2$\sigma$ $I$-band magnitude upper limit, $I_L < 21.77$, has
a probability of less than 0.5\% according to the discovery paper analysis. The reason for this discrepancy that \citet{dong-moa400} neglected the fact that much of the starlight in ground-based, seeing-limited images of the Galactic bulge contributes to the unresolved background in their
analysis when they estimated their upper limit on the lens star brightness. Thus, an upper limit
assuming that there is no starlight in the background can be overly strong.
This is a situation almost identical to the
overly strict lens star limits for MOA-2013-BLG-220L \citep{yee14}. The upper limits on the
brightness of the lens star for this event were also contradicted by the detection of the lens star
by \citet{van20}.

\section{Discussion and Conclusions}
\label{sec-discussion}
We have detected the planetary host star for microlensing event MOA-2007-BLG-400, and determined 
that the planetary lens system consists of a gas giant of slightly less than twice Jupiter's mass
orbiting a K-dwarf star. Our high angular resolution follow-up observations from Keck we have resolved the 
source and the lens to a separation of $\sim 96$ mas in the $K$ and $H$ bands, enabling us to accurately 
measure their fluxes and relative proper motion. We employed improved photometry methods to the majority
of the light curve data for this event \citep{bond17}, and this slightly decreased the Einstein radius crossing time, $t_E$, and 
mass ratio, $q$, parameters. Using this improved photometry and constraints from the Keck observations, 
we found that the host star is a K-dwarf located in, or very close to, the Galactic bulge. 
Due to the close-wide light curve degeneracy, we cannot be certain if the planet has a projected separation
of $a_{\perp} = 0.79\pm 0.10\,$AU or $a_{\perp} =5.9\pm 0.7\,$AU. The snowline of the host star is likely to be at 
$\sim 2.7 M_{\rm host}/\msun \simeq 1.86\,$AU \citep{kennedy_snowline}.
The close solution would likely put this gas giant planet well inside the snow line and
near the habitable zone of the host star, while the wide separation solution would put it
well beyond the snow line, where most microlens planets are found. There are very few microlensing planets 
detected so far inside the snowline of the host star, but because of the close-wide degeneracy we do not know
if the planet MOA-2007-BLG-400Lb lies inside or outside of the snowline.
\citet{suzuki16} shows the planet detection sensitivity for a mass ratio of 2.34$\times 10^{-3}$ of MOA survey spans 
a wide range of separations. The radial velocity technique has regularly detected Jovian planets just 
inside the snow line, so it would not be surprising if MOA-2007-BLG-400Lb was located either inside or well outside
of the snowline. In the future, if we know about planet occurrence rates at $\lesssim$ 1 AU from RV and transits, 
we can use this information to statistically break the close-wide degeneracy for such gas giant planets. 
That is, if we have a sample of close-wide degenerate events, we can use it to determine what percentage of the 
close-wide degenerate events are actually in the close configuration.

One of the most interesting features of Figure \ref{fig-lens} is that the host mass is much more massive than the
prediction from the Bayesian analysis that assumes that stars of all masses are equally likely to host planets. 
The measured mass is at the 93rd percentile of the predicted distribution shown as the blue histogram. 
This is nearly identical to the situation with MOA-2013-BLG-220L
\citep{van20}, which has a mass ratio of $q = (3.26\pm 0.04)\times 10^{-3}$, which compares to
the mass ratio of  $q = (2.34\pm 0.44)\times 10^{-3}$ for MOA-2007-BLG-400L. Both lens systems reside
in or very near to the Galactic bulge, and both are in the 93rd percentile of the 
predicted mass distribution, if we assume that all stars are equally likely to host planets of the observed 
mass ratios. This shows the importance of the mass measurements over the Bayesian mass estimates that are 
published in most of the microlensing planet discovery papers.  

\citet{laughlin04} have argued that the core accretion theory predicts that M-dwarfs are less likely to 
host gas giants than more massive stars. Earlier exoplanet results from radial velocities seem to 
support this idea \citep{johnson07,johnson10}, but these comparisons were at fixed planetary
mass, rather than fixed mass ratio. It would be a stronger statement to say that low mass stars are
less likely to host gas giant planets at a fixed mass ratio. However, the relatively large masses for the 
planetary host stars MOA-2007-BLG-400L and MOA-2013-BLG-220L might be an indication that 
M-dwarfs really are less likely to host gas giant planets (at a fixed mass ratio).
On the other hand, \citet{bennett20} found that planet OGLE-2005-BLG-071Lb is a $q = 7\times 10^{-3}$ 
planet orbiting an M-dwarf of $M = 0.43 \pm 0.04\msun$. This is a larger mass ratio, which might 
suggest that this planet might have been formed by gravitational instability \citep{boss1997,cameron1978} instead of core
accretion. It is also possible that the high metallicity of bulge stars plays a role \citep{bensby17},
because high metalicity stars have been found to be more likely to host gas giant planets \citep{fischer05}.
For events like OGLE-2005-BLG-071 \citep{bennett20}, OGLE-2005-BLG-169 \citep{batistaogle169}, 
MOA-2007-BLG-400, and MOA-2012-BLG-220 \citep{van20}, that have lens stars are now resolved
from their source stars, it should be possible to measure the metalicity. 
Of course, we do need a larger sample of microlens planets with host star mass measurements to resolve 
these questions, and this is the goal of our NASA Keck KSMS and HST observing programs.

This event is the first event to be published with lens detection in a microlensing system where the lens is $\sim$10 times fainter than the source star. It is much easier to detect the lens star when the lens and source are of comparable brightness and are separated by $\gtrsim$ 30 mas. This lens host is also one of the faintest host stars detected using Keck AO
\citep{blackman2020,terry2020}. 
% It is more difficult to detect these faint host stars since PSF variations play bigger role in uncertainties for faint stars. 
% Dave: the statement above is bullshit. The PSF variations are the same. You just need more exposure time to get a good measurement.
Our detection of the lens in a single high S/N image with a FWHM of $94\,$mas in 2013, when the lens-source separation
was $\sim 50\,$mas, implies that a full set of $\sim 20$ high S/N images would have enable the detection of the lens
star at a separation of $0.53\,$FWHM, with a contrast ratio of 10:1.
We estimate that if we managed to get more good S/N images in 2013, it would have been possible to detect the lens. 
It is clear that as the contrast between lens and source increases, it becomes increasingly harder to detect the lens. 
The {\it Roman} Galactic Exoplanet Survey will have $\sim$ 4.5 years difference between the first and the last 
epoch. Hence, the lens detection of an event with similar lens-source brightness contrast as MOA-2007-BLG-400 and 
a lens-source separation of $\sim$ 40 mas in 4.5 years might be challenging, if it were not for the much more
stable {\it Roman} PSFs and the many thousands of images that the {\it Roman} Galactic Exoplanet Survey will 
obtain for each event. However, events with much
smaller lens-source flux ratios might require follow-up observations with Extremely Large Telescopes equipped 
with Laser Guide Star AO systems and following the same image analysis methods discussed by
\citet{aparna19wp}. 

This planetary event is part of the \citet{suzuki16} statistical sample of 
30 planets found by the MOA microlensing survey and previous smaller statistical studies. 
This study and a follow-up analysis \citep{suzuki18} examine the exoplanet mass ratio function which is 
readily available from the light curve models,
but they do not examine how the exoplanet distribution depends on host mass. This event increases the number
of planets in this sample with mass measurements or upper limits to 12, and we have high angular resolution data for many
more of these events that are in the process of being analyzed. 
\citet{suzuki18} showed that the observational occurrence rate shows no deficit of intermediate mass giant planets,
with mass ratios in the range $10^{-4} < q < 4\times 10^{-4}$, and this contradicts a prediction based on the 
the runaway gas accretion process, which predicted a ``desert" in the distribution of exoplanets with these
mass ratios \citep{idalin04}. 3-D hydrodynamic simulations of gas giant planet formation seem to yield a similar
result \citep{szulagyi14,szulagyi16}. However, the \citet{suzuki16} does not include any dependence on
host star mass, and this is what we aim to correct with our high angular resolution follow-up observation program.
In particular, we plan to determine the exoplanet mass ratio function as a function of the host star mass, and to
investigate whether the exoplanet mass function provides a simpler description of exoplanet demographics
than the mass ratio function does.

We expect that high angular resolution follow-up observations combined with microlensing parallax measurements
will be able to measure or significantly constrain the host masses for 80\% of the \citet{suzuki16} sample, 
and we expect to expand this sample to include some planets recently found in the 2006 season data
\citep{bennett12,bennett_ob06284,kondo19}, as well as events that occurred more recently than 2012.

 This work made use of data from the Astro Data Lab at
NSF’s OIR Lab, which is operated by the Association
of Universities for Research in Astronomy (AURA), Inc.
under a cooperative agreement with the National Science Foundation. We also acknowledge the help of Dr. Peter Stetson on providing us with a feedback on our analysis of Keck data. The Keck Telescope observations and analysis
was supported by a NASA Keck PI Data Award 80NSSC18K0793. Data presented 
herein were obtained at the W. M. Keck Observatory from telescope time allocated to the National Aeronautics 
and Space Administration through the agency's scientific partnership with the California Institute of Technology 
and the University of California. The Observatory was made possible by the generous financial 
support of the W. M. Keck Foundation.
DPB and AB  
were also supported by NASA through grant NASA-80NSSC18K0274. This work was supported by the University of Tas-
mania through the UTAS Foundation and the endowed
Warren Chair in Astronomy and the ANR COLD-
WORLDS (ANR-18-CE31-0002). This research was
also supported in part by the Australian Government
through the Australian Research Council Discovery Pro-
gram (project number 200101909) grant awarded to
Cole and Beaulieu. Work by NK is supported by JSPS KAKENHI Grant Number
JP18J00897. AF's work was partly supported by JSPS KAKENHI Grant Number JP17H02871. 

%\begin{deluxetable}{cccc}
%\tablecaption{Measured Parameters from Single Star PSF fits \label{tab-single-fit}}
%\tablewidth{0pt}
%\tablehead{Passband&RA&Dec&Mag}
%\startdata
%HST $I$&&&\\
%HST $V$&&&\\
%Keck $K$&&&\\
%\enddata
%\end{deluxetable}

\end{document}